\documentclass{PoS}
\usepackage{epsfig}
\usepackage{graphicx}
\usepackage{longtable}

\newcommand{\solm}{M$_{\odot}$\ }

\title{The Multifrequency Behavior of Sagittarius~A*}
\ShortTitle{Sagittarius~A*}

\author{\speaker{Andreas Eckart}, M. Zajacek, M. Parsa, E. Hosseini\\
I. Physikalisches Institut der Universit\"at zu K\"oln, Z\"ulpicher Str. 77, D-50937 K\"oln, Germany;
and Max-Planck-Institut f\"ur Radioastronomie, Auf dem H\"ugel 69, D-53121 Bonn, Germany;
E-mail: \email{eckart@ph1.uni-koeln.de}
}
\author{
N. Fazeli, G. Busch, B. Shahzamanian, M. Subroweit, F. Peissker, N. Sabha, M. Valencia-S., M. Horrobin, C. Straubmeier, S. Rost, J. Schneeloch\\
I. Physikalisches Institut der Universit\"at zu K\"oln, Z\"ulpicher Str. 77, D-50937 K\"oln, Germany;
}
\author{A. Borkar, V. Karas\\
Astronomical Institute of the Academy of Sciences Prague, Bocni II 1401/1a, CZ-141 31 Praha 4, Czech Republic
}
\author{S. Britzen, A. Zensus\\
Max-Planck-Institut f\"ur Radioastronomie, Auf dem H\"ugel 69, D-53121 Bonn, Germany;
}

\abstract{
The Galactic Center is the closest galactic nucleus 
that allows us to determine the multi-frequency behavior of the supermassive black hole counterpart Sagittarius~A*
in great detail.
We put SgrA*, as a nucleus with weak activity, into the context of nearby low luminosity nuclei.
Possible hints for galaxy evolution of these sources  across the [NII]-based diagnostic diagram
can be inferred from dependencies on the masses, excitation ratios, and radio luminosities within this diagram.
For SgrA* we also discuss responsible radiation mechanisms 
covering results from the radio, infrared, and X-ray regime.
We also address the question of justifying the hot-spot model for describing flare profiles in light curves.
Since the brightness of LLAGN is also linked to star formation we briefly discuss the possibility of 
having stars formed in the immediate vicinity of supermassive black holes and possibly even in a mildly relativistic environment.
}
\FullConference{Multifrequency Behavior of \\
                High Energy Cosmic Sources - XII\\
		12-17 June 2017\\
		Mondello (Palermo), Italy}

\begin{document}


\section{Introduction}

Sagittarius A* (SgrA*) is located at the Galactic Center of the Milky Way.
It has been identified with an extremely underluminous, supermassive black hole (SMBH)
with a mass of 4 million solar masses and a quiescent bolometric luminosity 
of about 10$^{36}$ erg~s$^{-1}$. This is about eight orders of magnitude 
lower than the corresponding Eddington luminosity expected for such a black hole
(Narayan et al. 1998; Ghez et al. 2008, Eckart et al. 2017). 
It can, however, not be excluded that SgrA*
has been much brighter in the past, e.g., about 400 years ago 
(Revnivtsev Sunyaev \& Churazov 1998; Terrier et al. 2010, Witzel et al. 2012). 
SgrA* is the closest SMBH and therefore 
an ideal laboratory to study accretion and associated emission processes of SMBH systems 
(e.g. Falcke \& Markoff 2013).
As the origin of these flares two types of models are discussed:
electron acceleration processes
(Markoff et al. 2001; Liu et al. 2004; Yuan et al. 2004) and transient
accretion flows (e.g. Yusef-Zadeh et al. 2006ab, 2009; Eckart et al. 2006; Trap et al. 2011).
The emission mechanism is likely dominated by synchrotron and synchrotron self-Compton
processes. The exact importance of these processes is still under discussion.

Simultaneous multi-wavelength campaigns suggest that 
for the bright flares 
the radio, sub-mm, NIR and X-ray flare emission is correlated.
In this picture sub-mm flares follow concurrent NIR and X-ray out-bursts with time delays that depend on the observing frequencies below the 
sub-mm turnover frequency (Marrone et al. 2008; Trap et al. 2011; Eckart et al. 2012). 
These delays are probably governed by adiabatic expansion
of clouds of relativistic electrons (Subroweit et al. 2017, Yusef-Zadeh et al. 2006ab, 2008
Eckart et al. 2012, 2009; Marrone et al. 2008; van der Laan 1966). 
Flare emission of up to a few hours duration is due to non-thermal processes and 
can be studied from the radio to the X-ray domain (see references in e.g.  Mossoux et al. 2016, Eckart et al. 2017).
The quiescent X-ray emission is due to a thermal plasma extending out to the
Bondi radius (about 10$^5$ gravitational radii; e.g. Quataert 2002, Baganoff et al. 2001)

In the following we will discuss aspects of the multi-frequency behavior of SgrA*
covering results from the radio, infrared, and X-ray regime.
We will first compare SgrA*, as a weakly active nucleus, to nearby low luminosity nuclei (LLAGN)
and discuss possible hints for galaxy evolution of these LLAGN sources across the [NII]-based diagnostic diagram.
Since the brightness of LLAGN is also linked to starformation we briefly discuss the possibility of 
having stars formed in the immediate vicinity of supermassive black holes.
Specifically for SgrA* we also address the question of justifying the hot-spot model for describing flare profiles in light curves.

\begin{figure}[!ht]
\begin{center}
\includegraphics[width=13cm]{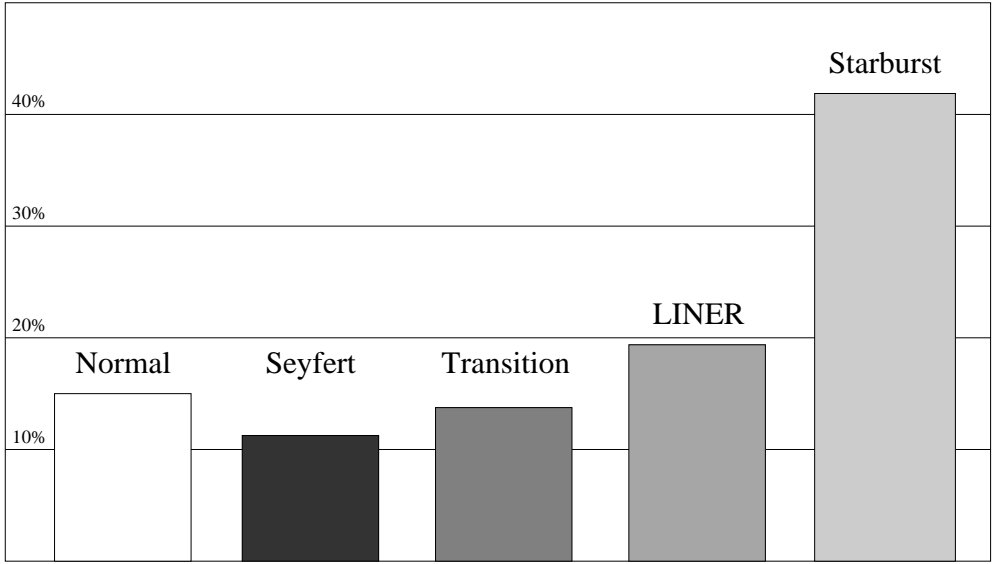}
\caption{
Demographics of activity in nearby galaxies
following Ptak (2001) and Ho, Filippenko, \& Sargent (1997).
} \label{fig1}
\end{center}
\end{figure}

\section{SgrA* as a Galactic Nucleus}

The presence of an Active Galactic Nucleus (AGN) or at least a supermassive black
hole at the center of a galaxy is a wide spread phenomenon.
As the Galactic Center provides highest angular resolution information of 
a galactic nucleus it is important to compare the source to other nuclei 
and to compare the different states of activities in these galactic cores.

\subsection{Low-luminosity AGN and normal galaxies}

\begin{figure}[!ht]
\begin{center}
\includegraphics[width=13cm]{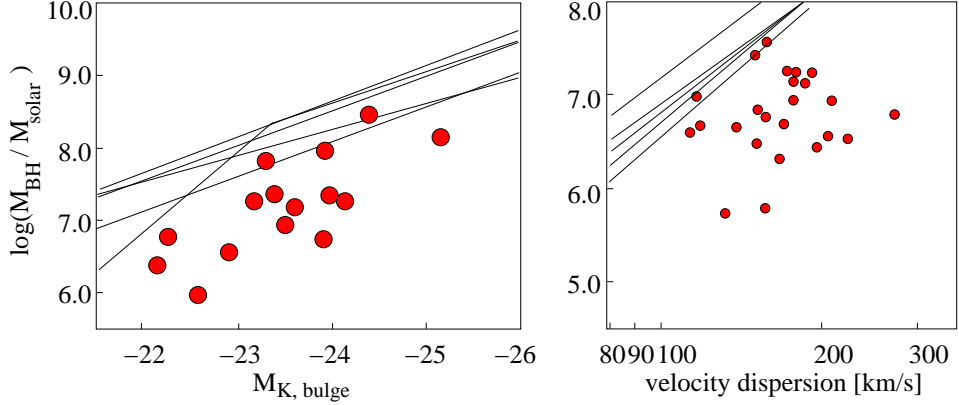}
\caption{
{\bf Left:} Correlation between black hole mass M$_{BH}$ and absolute K-band
magnitude M$_K$ of the spheroidal component as shown by Busch et al. 2016.
with black hole masses given in Busch et al. (2014, 2015, 2016).
{\bf Right:} The $M-\sigma$ plot with broad line AGN candidates as shown by
Subramanian et al. (2016).
Regression lines given by 
Ferrarese \& Merritt (2000),
Tremaine et al. (2002), 
Marconi \& Hunt (2003),
G\"ultekin et al. (2009), 
Vika et al.  (2012),
Kormendy \& Ho (2013), 
McConnell \& Ma 2013,
Graham \& Scott (2013)
(see details in Subramanian et al. 2016 and Busch et al. 2014).
} \label{fig2}
\end{center}
\end{figure}

Low-luminosity AGN with L$_X < 10^{42}$ ergs s$^{-1}$ are found much more 
frequently than ordinary AGN.  Therefore they may be more representative and 
also relevant to our understanding of the AGN phenomenon and the 
interplay between active nuclei and their host galaxies. 
Many normal galaxies are hosting LINER nuclei, starburst nuclei, or in general 
LLAGN (Low Luminosity AGN). Hence, they form the class of low-activity galaxies. 
These sources have rather similar X-ray characteristics, despite their 
apparently different optical classifications. 
This fact may be taken as an indication for a connection between 
starburst activity and AGN activity.
Following Ptak (2001) and Ho, Filippenko, \& Sargent (1997)
the demographics of activity in nearby galaxies is shown in Fig.~\ref{fig2}.

\begin{figure}[!ht]
\begin{center}
\includegraphics[width=13cm]{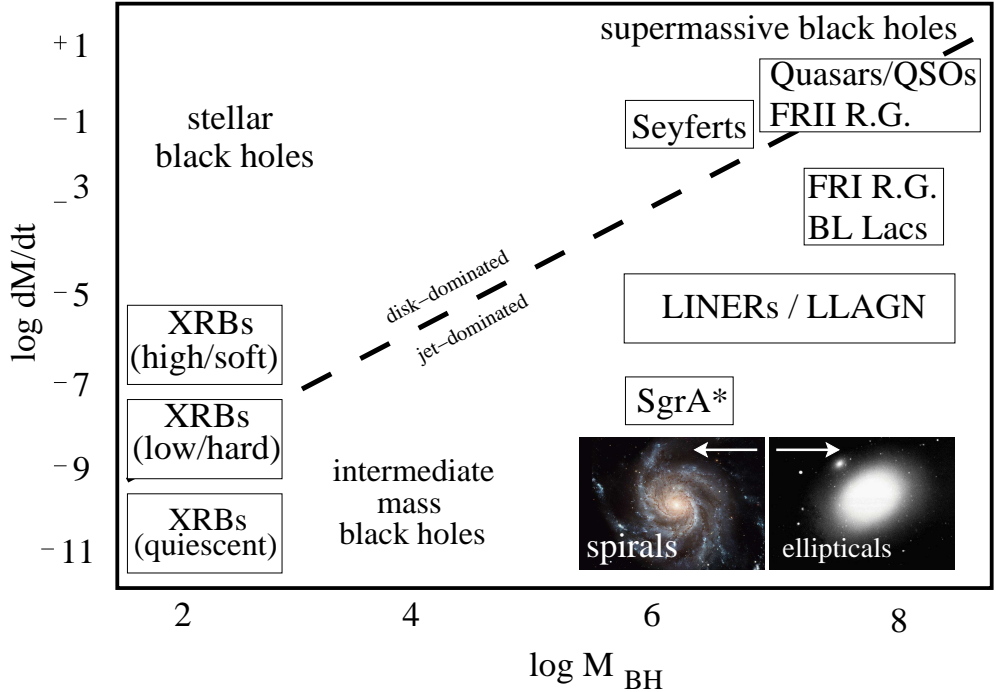}
\caption{
A proposed unification scheme for accreting black holes in the mass and accretion rate plane. Above a few percent of the Eddington accretion rate, the systems are proposed to be dominated by disk emission, while below they are inherently dominated by jet emission (RG = radio galaxy). Standard inclination-based unified schemes (Antonucci 1993; Urry \& Padovani 1995) are still assumed to be valid but are not explicitly shown here. Given a correlation between bulge mass and black hole mass, the AGN with the most massive black holes are supposed to reside in elliptical galaxies, while less massive black holes are predominantly in spirals. This is, of course, not applicable to XRBs.
} \label{fig3}
\end{center}
\end{figure}

The activity of the very compact active nucleus is most likely due to 
accretion onto a supermassive black hole. Tight correlations between the 
BH mass and host galaxy properties, particularly in early-type galaxies, 
are interpreted as signs for a co-evolution (see Kormendy \& Ho, 2013, 
for a review). There are, however, indications that the supermassive 
BH in LLAGN do not necessarily fall on the ``standard'' BH mass 
correlations for classical bulges and ellipticals (see below). 
Koliopanos et al. (2017) find that all LLAGN in their list have 
low-mass central black holes with $\log(M_\mathrm{BH}/M_\odot) = 6.5$ 
on average. They find that low surface-brightness AGN tend to have BH 
masses below the standard relations for spirals and ellipticals. 
Subramanian et al. (2016) investigate AGN black hole masses and 
the $M_\mathrm{BH}-\sigma$ relation for low surface-brightness galaxies. 
They compare them to existing $M_\mathrm{BH}-\sigma$ relations and find 
that they are mostly located in the low-$M_\mathrm{BH}$ regime and 
below existing $M_\mathrm{BH}-\sigma$ relations of inactive galaxies, 
which could indicate that they are not in co-evolution with their host galaxies
(see Fig.~\ref{fig3}, right).
In Busch et al. (2016a), we determined BH masses for 16 AGN taken from 
the low-luminosity type-1 QSO sample of the 99 closest ($z\leq 0.06$) 
QSOs from the Hamburg/ESO survey for bright UV-excess QSOs. 
The sources show an offset from the classical 
$M_\mathrm{BH}-L_\mathrm{bulge}$ relations
(see Fig.~\ref{fig3}, left).
But they show signs of nuclear star formation activity (Busch et al., 2015), 
which suggests that this offset may be due to an over-luminosity of the bulges.
A further analysis promises important insight in the evolution scenarios 
of supermassive black holes and their surrounding host galaxies (Busch 2016b).

Falcke, K\"ording \& Markoff (2004) use a
scheme of an accretion rate-dependent transition between states
to explore the evolution in power and the presence of jets for 
stellar to supermassive black holes  (see Fig.~\ref{fig3}).
For black hole systems that accrete near the Eddington rate 
the emission will be dominated by the accretion disk. 
As a common feature the assumption is that below a 
critical accretion rate value the radiation and accretion mode 
becomes optically thin as well as radiatively inefficient. 
If that is the case, the emission of these sub-Eddington systems
will be dominated by synchrotron emission from a relativistic jet.
These sub-Eddington systems can be found for all black hole masses.
Amongst them is 
the Galactic Center (SgrA*), but also 
low-ionization nuclear emission-line region galaxies (LINERs; Heckman 1980ab),
Fanaroff Riley type I (FR~I) galaxies,
and BL Lacertae (BL~Lac) objects. 
There also appear to exist scaling relations between stellar 
(e.g. in X-ray binaries) and supermassive black  hole systems.
Falcke, K\"ording \& Markoff (2004) place black holes of all masses into
a scheme that depends mainly on the black hole mass, the accretion rate, 
as well as the relative importance of disk and jet structures.

\subsection{Hints for evolution}
\label{sec:Hintsforevolution}

It is currently unclear what role (if any at all) active galactic nuclei (AGN) play with respect to galaxy evolution. 
Without doubt there is a connection between the central black holes and the bulges in which they are found.
In addition to density dependent stellar dynamical effects (like collisional interaction between stars and core collapses of dense
stellar systems)
AGN feedback is one of the most promising candidates to connect the AGN with the host galaxy, the ISM, and star formation
processes that are taking place in these galaxies.
However, these influences may also be connected to or covered up by galaxy-galaxy interactions like 
harassment, collisions and subsequent mergers.

The detected AGN in a combined optical-radio sample show a shift of the AGN region of the diagnostic diagram with increasing 
L$_{20~cm}$/L$_{H \alpha}$. 
The diagnostic diagrams show enhanced starforming or composite region ratios with log(L$_{20~cm}$/L$_{H \alpha}$)$<$0.716.
For LINER regions this enhancement peaks above this value. 
Vitale et al. (2012) show that a comparison with photo-ionization models and shock models indicates that the LINER
phenomenon may be connected to enhanced importance of shocks in and around these nuclei.
Our results indicate that it is worthwhile to include the radio properties in the classification scheme in order to 
better differentiate between Seyferts and LINER sources.

\begin{figure}[!ht]
\begin{center}
\includegraphics[width=13cm]{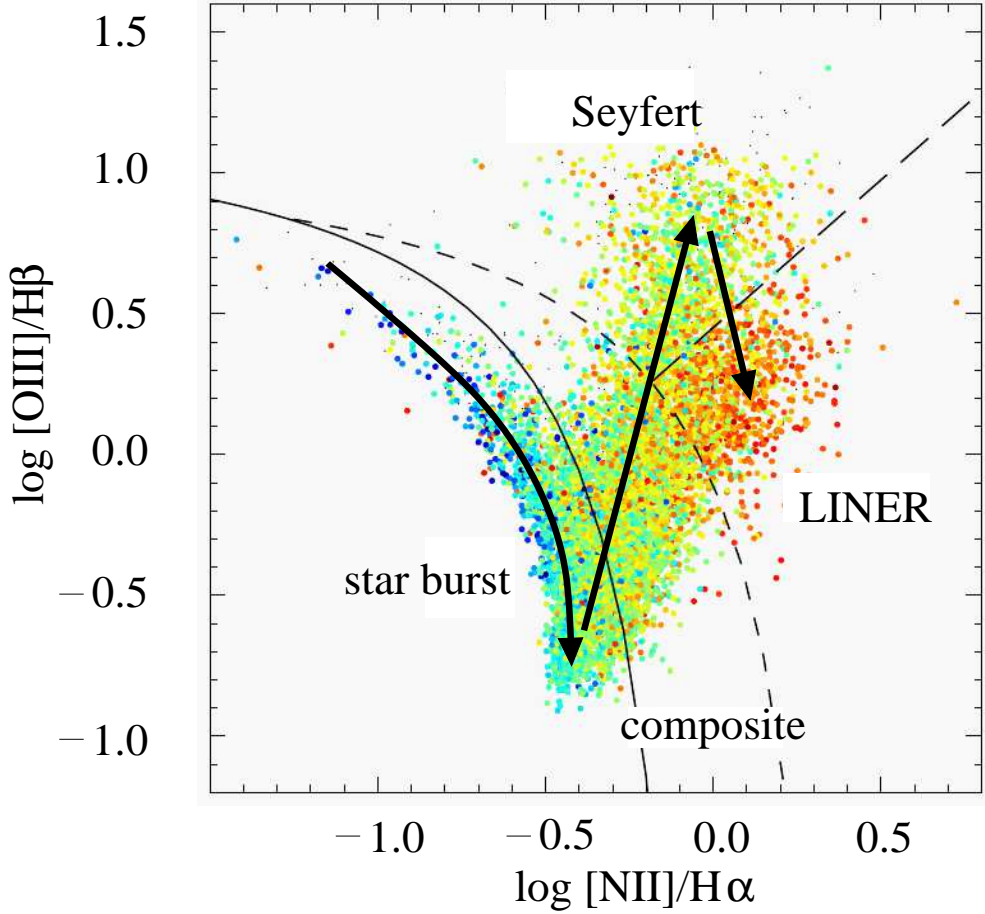}
\caption{
Sketch of galaxy evolution across the [NII]-based diagnostic diagram
based on figures by Vitale et al. (2012). 
The arrows represent the trend of possible galaxy evolution from starforming galaxies to Seyferts and LINERs.
The colors indicate the stellar host mass in solar mass units
blue: 10$^{9}$ to 10$^{9.8}$ \solm;
cyan 10$^{9.8}$ to 10$^{10.3}$ \solm;
green 10$^{10.3}$ to 10$^{10.8}$ \solm;
yellow 10$^{10.8}$ to 10$^{11.3}$ \solm;
red 10$^{11.3}$ to 10$^{12.1}$ \solm.
} \label{fig4}
\end{center}
\end{figure}

Studies of the AGN populations in the radio-optical domain are therefore crucial
if one aims at understanding the importance of AGN feedback. 
In particular  a mass sequence linking the starforming galaxies and the AGN activity has been recently discussed.
These processes and dependencies may point at possible evolutionary sequences.
As highlighted by Vitale et al. (2015) we observed a sample of 119 intermediate-redshift 
SDSS-FIRST radio emitters in a redshift interval between z=0.04 and z=0.4 using the Effelsberg 100-m telescope.
From observations at 4.85 and 10.45 GHz we obtained flux densities and spectral indices. 
The sample covers starforming galaxies, composite galaxies (with a line emission contribution from star formation and AGN activity), 
Seyferts, and low ionization narrow emission region (LINER) galaxies 
(Fig.~\ref{fig4}).
The results are discussed in a [NII]-based emission-line diagnostic diagram. We find that the radio spectral index 
flattens along the increased host and nuclear black hole mass from high-metallicity starforming galaxies via composite galaxies to 
Seyferts. Hence, the flattening goes along with a hardening of the ionizing field as indicated by the line ratios in the
diagnostic diagram. There are also systematic variations in the spectral curvature and the radio structure (i.e. the 
relative contributions of cores and jet/lobes).
These findings by Vitale et al. (2015) are a clear indication for the fact that the galaxies along this sequence are 
in a transition from star forming galaxies, via a nuclear radio phase, to more  passive elliptical galaxies. 
(Fig.~\ref{fig4}).
Hence, within this picture it is also supported that the AGN feedback is playing a role in shutting down star formation 
activity.

The interplay between black hole masses, multifrequency luminosity 
of the nucleus, and starformation, i.e. a feedback between the nucleus and its immediate vicinity,
may also be reflected in the properties of the water maser disks
found in the nuclei of these galaxies.
Through VLA radio continuum observations 
Kamali et al. (2017) investigate the nuclear environment of galaxies 
that have subparsec edge-on accretion disks
with observed 22~GHz water megamaser. 
The black hole masses show stronger 
correlations with water maser luminosity than with 1.4 GHz, 
33~GHz, or hard X-ray luminosities. 
Furthermore, the inner radii of the water maser disks show stronger 
correlations with 1.4~GHz, 33~GHz, and hard X-ray luminosities 
than their outer radii, suggesting that the outer radii 
may be affected by disk warping, star formation, 
or peculiar density distributions.

Independent of stellar processes like star formation, a significant part of all nearby galaxies 
shows some evidence for a weak AGN (e.g. Ho et al. 2008, Eckart et al. 2012).
That trend becomes clear with increasing angular resolution and an increased coverage of the 
overall electromagnetic spectrum.
The essential ingredients one needs to set up of specify nuclear activity in the vicinity of a 
supermassive black hole are 
(1) a compact radiatively inefficient accretion flow, 
(2) a possibly truncated thin accretion disk, and 
(3) a jet or some loosely collimated outflow. 
Detailed investigations of nearby galaxies, AGN and Low Luminosity AGN 
indicate that many, maybe all, 
galactic bulges host a central supermassive black hole. The exception may be dwarf galaxies and some
late-type galaxies. 
This underlines the importance of a deeper understanding of LLAGN, 
to which studies of SgrA*, the nearest galactic nucleus which also 
shares many properties of typical LLAGN, can give important contributions.

\subsection{Hints for relativistic effects near SgrA*}
\label{sec:Hintsforrelativity}
\begin{figure}[!ht]
\begin{center}
\includegraphics[width=13cm]{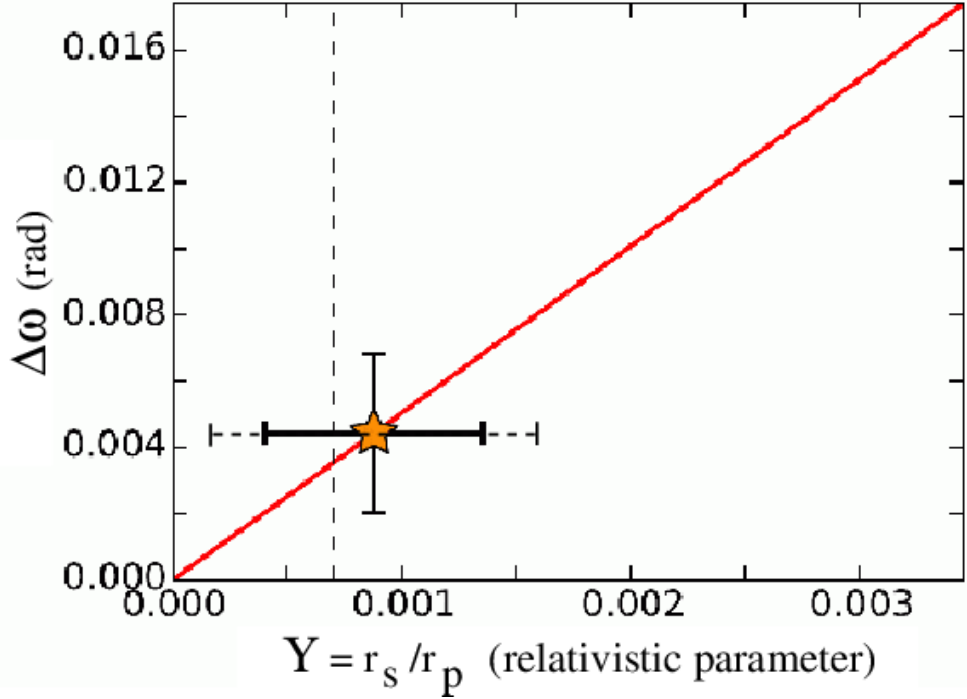}
\caption{
Correlation between the relativistic parameter $Υ$. 
and the periapse shift $\Delta$$\omega$. 
S2 is represented by an orange asterisk symbol. 
The long vertical dashed line is located at the expected value for
$Y$. The solid lines indicated the median and the absolute median deviation
derived for $Y$ and $\Delta$$\omega$ using the orbital data from S2.
The dashed horizontal black line marks the median deviation derived from
the measured  $\Delta$$\omega$ and the measured relativistic variations 
in the large half axis and the elipticity of the S2 orbit 
(see Parsa et al. 2017 for more details).
} \label{fig5}
\end{center}
\end{figure}

In order to properly interpret the multi-frequency behavior of SgrA* it is measuring 
to verify that the large mass associated with SgrA* has the influence on space time as expected.
Parsa et al. (2017, Fig.~\ref{fig5}) recently suggested  
that the young S-cluster stars are indeed experiencing relativistic effects due to the 
large central mass.
Here, the authors primarily aimed at a new method to investigate relativistic orbits of 
stars in a strong gravitational field like we have it close to SgrA*. 
The authors used a first-order post-Newtonian approximation and calculated stellar orbits with a 
wide range of impact parameters, i.e. periapse distances to the supermassive 
black hole $r_p$ (here taken as an impact parameter). 
Close to a supermassive black hole the orbits start being affected by relativity effects
which leads to characteristic changes in the orbital elements.

The significance with which the relativistic environment can be proven is currently limited:
Studies of the robustness of the result by Parsa et al. (2017) show that in case of strong noise
a positive result would indicate at least a 3-4~$\sigma$ detection assuming
that the measurement uncertainties follow a Gaussian process.
A positive result means: deformations of the orbital shape that are in quality
consistent with the expected deformations due to the influence of relativity.
Such a result is found by Parsa et al. (2017).
The actual magnitude of the effect (i.e. by how
much the correct distortion of the shape is realized) has been calculated
by tracing the stellar position measurements plus their uncertainties
through the reduction and analysis pipeline.
The actual values obtained by Parsa et al. (2017) that describe the orbital
deformation are in good agreement with recently predicted values from theory.
Here, Iorio et al. (2017) studies Post-Keplerian effects on radial velocity
in binary systems and the possibility of measuring the effects of
General Relativity with the Galactic Center star S2 in 2018.

These changes are also correlated with parameters that characterize the degree of relativity
present in the distance range covered by the stellar orbits.
To express the result though a parameter that measures the degree of relativity
that depends on quantities that measure the corresponding distortions
Parsa et al. (2017) had to describe the results using truncated normal
distributions a method that is often used in statistics and econometrics.
It is used if dependent variables can take only a limited number or a limited range of values.
The combination of different measures of the distortion does not follow a Gaussian distribution.
Hence, Parsa et al. (2017) then used for the final derivation of the relativistic parameter
the median and the median deviation rather than the mean and the standard deviation.
One of these relativistic parameters can be defined as the ratio between the  Schwarzschild radius
 $r_s$ and the impact parameter $r_p$. This ratio 
${\Upsilon} = r_s/r_p$ can be obtained from observational data. 
For the star S2 Parsa et al. (2017) find a value of ${\Upsilon} = 0.00088 \pm 0.00080$.
the expectation for S2 is 
${\Upsilon} = 0.00065$, as it can be derived from the SgrA* mass and the orbit of S2 approximated 
as a Keplerian orbit.
While the results can certainly be improved through the upcoming
interferometric measurements in the infrared, it is the first time that a relativistic 
parameter could be obtained from stellar observational data.
The currently obtained value is unlikely to be 
dominated by perturbing effects like excess noise on the stellar 
positions, rotation of the field, drifts in black hole mass, or perturbing influences of 
other masses within the central S-cluster (see discussion by Parsa et al. 2017). 

Comparison to Mercury:
The relativistic orbit deformations expected and indicated for the star S2
orbiting the 4.3 million solar masses black hole at the center of the Milky Way
are indeed the comparable to those detected for the orbit of Mercury around the Sun.
There is, however, one difference! The relativistic Mercury pericenter shift
is only about 1/12 of the shift that Mercury's orbit experiences from the
forces by the other planets: It is 43 arcseconds out of 532 arcseconds.
For the star S2 the interaction with the other stars on a single orbital time scale
is much smaller. Depending in the presence of 20 to 30 solar mass black holes the
effect is either comparable to or (in case of the absence of such massive scatterers)
almost negligible compared to the relativistic effect due to the 4.3 million solar masses of SgrA*
(Sabha et al. 2012).
For pure, undisturbed relativistic effects,
we expect to see for star S2 a pericenter shift of 11 arcminutes (i.e. 660 arcseconds) and we measured
14$\pm$7 arcminutes (i.e. 840$\pm$420 arcseconds). So - if the measurement can be
made more precise in the near future, S2 provides a much clearer case for the
relativistic pericenter shift when compared to Mercury.

Precise positional measurements are the key to further progress
in proving SgrA*'s relativistic environmemt.
Larger telescopes are needed to achieve that. One way of getting to larger telescopes
is interfering their received radiation at a single combined focus.
GRAVITY is an interferometric beam combiner that allows one to connect the four 8.2 meter diameter
unit telescopes  of the European Southern Observatory on the Paranal mountain in Chile.
This way a 120 m diameter telescope can be realized.
Repeated measurements of the position of the NIR counterpart of the SgrA* black hole
and the star S2 will allow an exact measurement of the S2 orbit and
a much more precise determination of the associated relativistic effects.

\begin{figure}[!ht]
\begin{center}
\includegraphics[width=13cm]{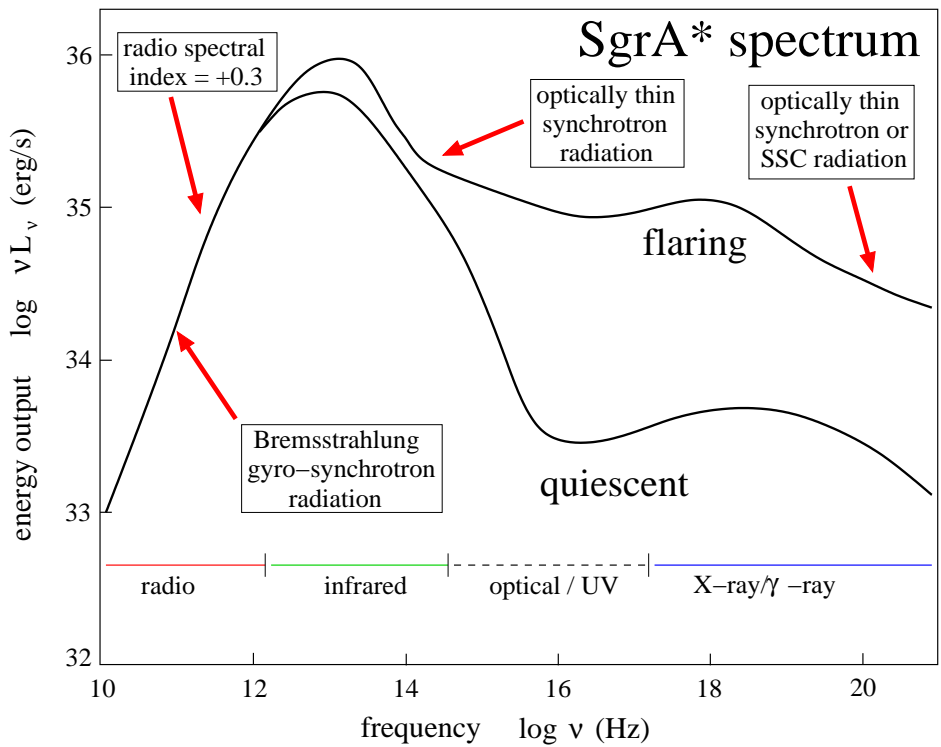}
\caption{
Spectrum of SgrA* in quiescent and flaring state
(see e.g. Yuan et al. 2003, Narayan et al. 1998, 2008).
} \label{fig6}
\end{center}
\end{figure}

\subsection{Hints for starformation in a mildly relativistic environment}
\label{sec:Hintsforstarformation}

Currently the most general simultaneously obtained estimates for the mass of and the distance to 
the supermassive black hole SgrA* involve the three stars that have the the shortest period 
(i.e. S2, S38, and S55 - alias S0-102) based on Newtonian models are 
M$_{BH} = (4.15\pm0.13\pm0.57) \times 10^6$ \solm and $R_0 = 8.19\pm0.11\pm0.34$kpc 
(Parsa et al. 2017). 
The presence of this large mass as well as the presence of young stars in the central
stellar cluster indicates that stars are being formed (even in the mildly relativistic)
in the immediate vicinity of supermassive black holes.
The young ages of only a few million years as well as their
low orbital eccentricities 
indicate that the formation of the more massive emission line stars probably 
took place in-situ in a dense accretion disk of gas and dust
(Nayakshin \& Cuadra 2005, Levin 2007; Alexander et al. 2008; Bonnell \& Rice 2008).
In Jalali et al. (2014) we showed how ongoing starformation 
can be sustained in the vicinity of a very large black hole mass. 
This may include also of stars like the high velocity S-stars that have masses 
between 8 and 14 \solm (Habibi et al. 2017) or even lower mass stars similar to the
Dusty S-cluster Object (DSO) that recently went through a SgrA* periapse
(e.g. Zajacek et al. 2014, 2016, 2017, Valencia-S. et al. 2015).
Jalali et al. (2014) show that infalling gaseous clumps 
(possibly from the circum  nuclear ring) can be 
further compressed during periapse (if sufficiently cold i.e. less than 50 to 100~K) such that 
the formation of 
main sequence stars is in fact enhanced due to the presence of a large mass. 
The X-ray footprint (i.e. shadow agains the local Galactic Center X-ray background) of the Circum Nuclear Disk
has recently been revealed by Mossoux \& Eckart (2017). 
While this phenomenon indicates that star formation is actually triggered in a mildly relativistic 
environment this needs to be further supported. 
The S-star cluster in the Galactic center is a particularly well suited place to study the 
physical processes in the vicinity of a supermassive black hole.
This is also the region where stars can be used to perform dynamical tests of general relativity. 

\section{Radiation mechanism}
\label{sec:Radiationmechanism}

\subsection{Overall Spectrum}
\label{subsec:Radiationmechanism}

In Fig.~\ref{fig6} we show a sketch of the quiescent and flaring 
broad band spectrum of SgrA* based on radio data 
from Falcke et al. (1998) and Zhao et al. (2003), 
the IR data from the VLT (see e.g. Eckart et al. 2017 and references in there).
The X-ray date are from e.g. Baganoff et al. (2001, 2003),
Porquet et al. (2003, 2008) and Neilsen et al. (2013).
The solid lines are based on ADAF models by
e.g. Yuan et al. (2003), Narayan et al. (1998, 2008).
The nonthermal variable source associated with SgrA* can only be 
observed in the radio to longer mid-infrared domain
(see e.g. Sch\"odel et al. 2011).
During the quiescent phases it is currently below the detection/confusion
limit in the NIR to X-ray domain.
During its flare phase it is well visible in all spectral domains
including harder X-rays (see e.g. Zhang et al. 2017).
In Fig.~\ref{fig6} we have highlighted some of the prominent 
radiation mechanisms from gyro-synchrotron, Bremsstrahlung, synchrotron 
and synchrotron self-Compton (SSC) radiation.

\subsection{Flares}
\label{subsec:Radiationmechanism}

While the flare emission clearly indicates that some accretion, 
heating, or magnetic acceleration mechanism give rise to enhanced
emission, one has to take into account the fact that this emission 
arises in the immediate vicinity of a supermassive black hole,
i.e. a large mass with relativistics effects as a consequence.
General Relativistic Magneto Hydrodynamics (GRMHD) allows to understand
important basics of the emitting region around SgrA* including
mid-planes or disks, jets, outflows and more extended emission zones
(e.g. Moscibrodzka et al. 2017, 2014, 2013, Broderick et al., 2011,
Shcherbakov, Penna, \& McKinney 2012).
Short term variability is often explained by hotspot emission from the 
mid-plane that often forms in GRMHD simulations due to the conservation 
of angular moment (e.g. Meyer et al. 2006ab, Eckart et al. 2006, 
Broderick \& Loeb, 2005, 2006, Bao \& Ostgaard 1995, Bambi 2015).
However, close to the black hole relativistic lensing and boosting effects
influence the appearance of flare emission
(e.g.  Broderick \& Loeb, 2005, 2006, Eckart et al. 2006).
In some bright X-ray flares these characteristic shapes may be visible
(see \ref{subsubsec:Xray}).

Based on the hotspot model, Karssen et al. (2017) address if the observed
light curves of X-ray flares can give independent constraints on the mass
of the central supermassive black hole. They study
the asymmetric flare shapes that arise from temporary flare of a blob
orbiting near a supermassive black hole.
The inferred mass derived from the flares is in agreement with previous mass
estimates based on orbits of stars (e.g. Parsa et al. 2017, and references therein).
Karssen et al. (2017) also successfully test the reliability of the method by applying
it to lightcurves from the Seyfert~I galaxy RE~J1034+396.

These characteristic flare shapes are a result of an interplay of
relativistic effects that are responsible for the modulation of the observed light curves:
Doppler boosting, gravitational redshift, light focusing, and light-travel time delays.
In order to see these shapes it is required that the intrinsic hotspot emission
stays rather constant during the phase in which the important relativistic 
amplification effects can take place.
In the following we discuss the flare emission mechanisms
under this aspect.
The goal, here, is to narrow down the relevant time scales on which the 
intrinsic flux density of the source component
varies before experiencing rapid and strong modulations by relativistic 
effects like Doppler-boosting (beaming) or lensing.

In Fig.\ref{fig8} we show data of two bright flares observed with the
Chandra/HETGS  instrument in addition with a hotspot flare 
model with parmeters given in the caption.
The shoulder is due to lensing when the hotsopt is behind the black hole
and the peak is due to boosting when the hotspot approaches the observer
after passing around the backside of the black hole.

Since bright X-ray flares are often seen synchronous with brigh NIR-flares
the finding by Karssen et al. (2017) is probably also applicable 
for NIR-flares.
The method by Karssen et al. (2017) indicates that at least for bright flares 
the flare time scales are coupled to the dynamical time scale of radiation 
matter colse to SgrA*.

\subsubsection{The near-infrared domain}
\label{subsubsec:NIR}
The value of the spectral index for brighter flares 
which gives us information on the relativistic electron energy distribution
is consistent with $\alpha_{NIR/MIR}$=-0.7 (S$\sim$$\nu^{+\alpha}$).
This is in agreement with optically thin NIR synchrotron radiation.
In Bremer et al. (2011; see also references there in) 
we highlight a possible tendency for weaker flares to exhibit 
a steeper spectrum. We conclude that the distribution of spectral indices
as a function of NIR K-band flux density can successfully be described by an exponential cutoff proportional to
exp[-($\nu$/$\nu_0$)$^{0.5}$] because of synchrotron losses, with $\nu_0$ being a characteristic cutoff frequency. 
Varying $\nu_0$ between the NIR and sub-mm domain and assuming a sub-mm flux density variation of about one Jansky and optically
thin spectral indices of $\alpha_{NIR/MIR}$=-0.7$\pm$0.3 explain the observed spectral properties of SgrA* in the NIR (Bremer et al. 2011).
The infrared 'flares' are bright flux density excursions of
the variable NIR emission that can be described well by a stationary red-noise flux density distribution of
the form  $p(x)\propto x^{-\beta}$ with a power-law index $\beta_{NIR}\sim4$
(Witzel et al. 2012).

\begin{figure}[!ht]
\begin{center}
\includegraphics[width=13cm]{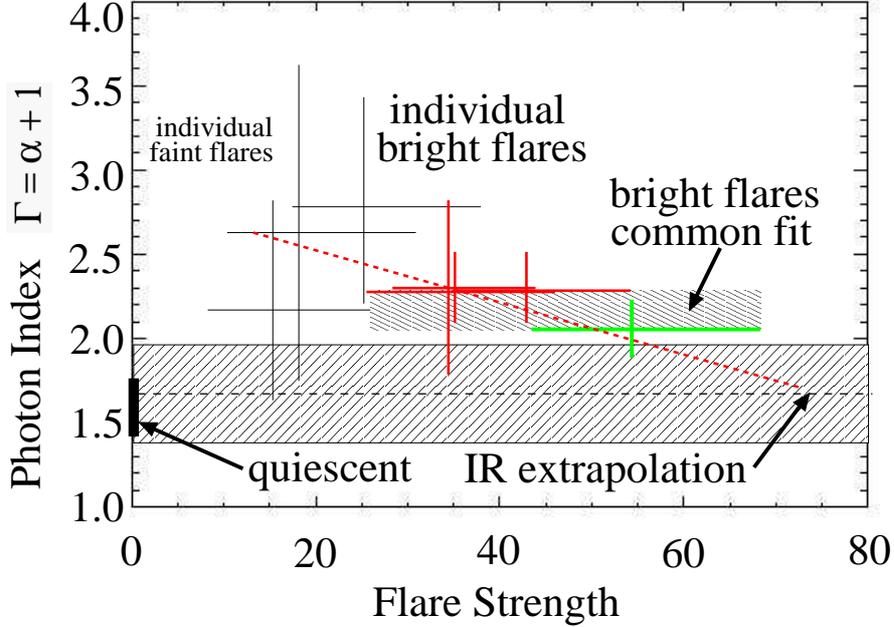}
\caption{
Photon index $\Gamma$ versus flare strength for flares in the X-ray domain. 
The flare strength is given in multiples of the quiescent state flux from 
the X-ray source at the position of the radio source SgrA*
with S$\sim$$\nu^{+\alpha} =\nu^{+\Gamma-1}$ and  $\Gamma = \alpha +1$.
Individual bright flares are shown in red and green, individual faint
flares are shown in black error bars.
The red dashed lone shows a tendency for flattening towards high flare fluxes
(see also line in Fig.3 by Zhang et al. 2017).
} \label{fig7}
\end{center}
\end{figure}

\subsubsection{The radio domain}
\label{subsubsec:radio}
Observational evidences and theoretical modelling show that the 
synchrotron spectra which are optically thin in the NIR peak 
in the 300-500~GHz band 
(Eckart et al. 2008, 2012, Marrone et al. 2006ab, 2006PhDT).
In Subroweit et al. (2017) we perform a statistical analysis of both the variable 100 and 345~GHz flux densities of SgrA*.
We find that both distributions are also well described by power-laws with $\beta_{NIR}$$\sim$4. 
Within the framework of a plasmon model we also use the derived power-law indices to constrain the important model
parameters: We find the initial synchrotron turnover frequency $\nu_0$ and  the source component expansion velocity 
$v_{exp}$ to be predominantly above 100~GHz and below 0.01~c, respectively (see details in Eckart et al. 2012).

\subsubsection{The X-ray domain}
\label{subsubsec:Xray}
Neilsen et al. (2015) presented a statistical analysis of the X-ray flare flux distribution. 
They also described the flux counts by a power-law distribution with $\beta_{X-ray}$$\sim$2. 
Neilsen et al. (2015) point out that in case of the X-ray emission being produced via the 
Synchrotron Self Compton (SSC) effect, this 
index is fully consistent with the power-law index $\beta_{NIR}\sim4$ (Witzel et al. 2012).
Hence, it is very likely that the X-ray flare emission is fully dominated by SSC radiation.
This is in agreement with the modeling of 8 flares with simultaneous NIR- and X-ray measurements
by Eckart et al. (2012).
For these flares the caculations indicate that for synchrotron flares in the infrared with the
corresponding SSC response in the X-ray domain the volume density of relativistic electrons
is larger than 10$^{6}$cm$^{-3}$ with a typical expectation value of $\sim10^{9.5}$cm$^{-3}$
high but not excessive (Eckart et al. 2012 and Mossoux et al. 2016).

If one wants to explain the X-ray flare
emission by synchrotron radiation only, then the high energy cutoff in
the electron energy distribution requires large Lorentz factors  of $\gamma_e > 10^5$ 
for the emitting electrons and magnetic field strengths
in the range of 10-100~G (Baganoff et al. 2001, Markoff et al. 2001, Yuan et al. 2004). 
However, if $\gamma_e > 10^5$
electrons are being produced it is very likely that a lot more 
$\gamma_e > 10^3$ electrons are available as well. Hence, the SSC emission by those lower
energy electrons will already produce a significant portion of the 
observable X-ray emission 
such that it becomes hard to justify 
 the need for a significant population of $\gamma_e > 10^5$
electrons (see discussion by Eckart et al. 2006, 2008, 2012).

Zhang et al. (2017) show in their Fig.4 a decomposition of the NuStar X-ray spectrum of SgrA*
as obtained for the inner 30 arcseconds of the Sgr A* emission complex during its X-ray quiescence.
They show the data together with the best-fit model. 
Several components are indicated: two thermal components with kT$\sim$1.1~keV and kT$\sim$6.7~keV, a
Gaussian model to fit the 6.4~keV neutral Fe-line, and a power-law (blue).
The thermal components become very faint above about 20~keV. Then the non-thermal power-law 
component starts to dominate the spectrum. 
This spectrum may constrain the SgrA* quiescent luminosity level, however it may also 
contain contributions from a Pulsar Wind Nebula (PWN) candidate and the diffuse X-ray emission
in that region.

In Fig.\ref{fig7} we show the 
photon indices versus flare strength for 7 flares
detected by NuStar with significance of more than $>5\sigma$.
The flare strength is defined as the ratio of the flare 2-10~keV 
unabsorbed flux to the quiescent state flux of
F(2-10~keV) = $0.47_{-0.03}^{+0.04}\times10^{-12} erg~cm^{-2}s^{-1}$.
There is no significant correlation between the flare spectral shape and the
flare luminosity.
For comparison we also show the photon index for the 
quiescent state (thick bar to the left), the common fit to the three bright 
NuStar detected flares (shaded box), and the expected photon index using
the spectral index from the near-infrared domain (light shaded box).
This NIR spectral index would be valid if the spectrum is due to 
synchrotron radiation and the spectrum extends into the X-ray domain
or if the spectrum is due to SSC radiation staying with the same spectral index 
as the synchrotron spectrum it originates from.
There is the tendency that this may indeed be the case for the brightest flares
(see dashed line in Fig.\ref{fig7} and in Fig.3 by Zhang et al. 2017).

\begin{figure}[!ht]
\begin{center}
\includegraphics[width=13cm]{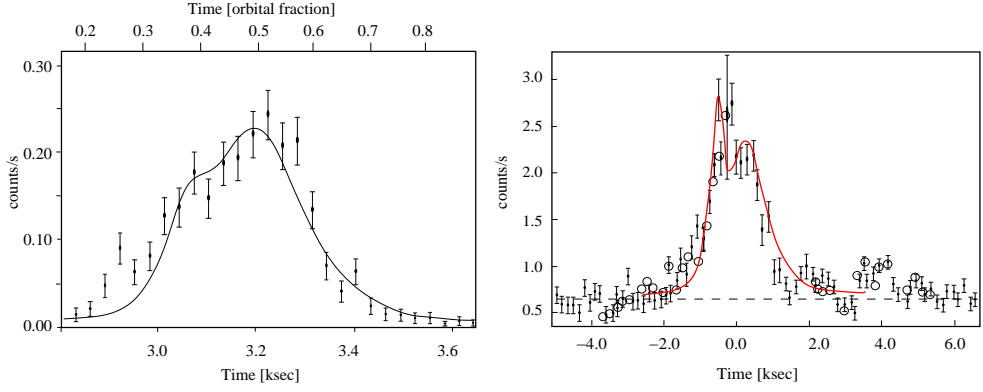}
\caption{
{\bf Left:} Bright X-ray flare published by Nowak et al. (2012)
shown with hotspot model fit by Karssen et al. (2017) for a 
spin of a=0.5,
an inclination of 80$^o$,
a spot radius of 18~$r_s$,
a spot size of 5~$r_s$.
{\bf Right:} The bright flare reported by Ponti et al. (2017)
compared to a hotspot model for a
spin of a=0.5,
an inclination of 90$^o$,
a spot radius of 18~$r_s$,
and a spot size of 2.5~$r_s$.
In both cases the black hole mass implied by the X-ray flares is about 3.5$\times$10$^6$\solm
and compares well with the current value of
M$_{BH} = (4.15\pm0.13\pm0.57) \times 10^6$ \solm 
derived form stellar dynamics
(Parsa et al. 2017). 
} \label{fig8}
\end{center}
\end{figure}

\subsubsection{Cooling time scales in the X-ray domain}
\label{subsubsec:cooling}
Cooling time scales for the synchrotron components peaking in the sub-mm are of the order of
1-2 hours (see also Eckart et al. 2006).
In this frequency range the space density of photons 
scattered by the cospatially present relativistic electrons is highest.
This implies that the intrinsic flare light curve in the X-ray domain 
is dominated through the SSC process by the cooling time scales of the sub-mm synchrotron components.
This is contrasted by the expectations for direct synchrotron production of the X-ray emission.
In this case the high energy electrons result in short cooling time scales of less than a few hundred seconds.
Source variability on these time scales would be expected and longer
flares would require repeated injections or acceleration of high
energy electrons (Baganoff et al. 2001, Markoff et al. 2001, Yuan et al. 2004, Eckart et al. 2006, 2012). 

However, the observational facts described above make it more likely that
a population with energetically far less demanding $\gamma_e \ge 10^3$ electrons 
and an upper synchrotron cutoff frequency just short of the NIR
provide the dominant portion of the X-ray emission through the SSC process.
Such a scenario is also greatly supported by the close statistical relation between the 
flare flux power-law indices $\beta_{NIR}$ and $\beta_{X-ray}$ (Neilsen et al. 2015)
and the modelling of individual flares (Eckart et al. 2012).
The sub-mm cooling time scales (and adiabatic expansion time scales for the spectral peak emission 
from the sub-mm to the radio) are also of the same order as the length of the bright X-ray flares. 

As the sub-mm to NIR domain is linked to the X-ray domain via SSC scattering, one may also 
look at the time scales associated with quantities that dominate the scattering efficiency
(see equation 4 by Marscher 1983):
Quantities like the source size $\theta$ and the synchrotron
turnover frequency $\nu_m$ are governed by the time scales of adiabatic expansion (hours).
Quantities like the synchrotron peak flux $S_m$  and the spectral 
index $\alpha$ are governed by the accretion process. 
Here, for SgrA* the times scales go from 
the orbital time scale (i.e. typical flare time scales) to many days  
(specifically for X-ray variability see equation 5 by Ishibashi et al. 2012).
Also shocks that may influence the relativistic particle distribution 
or the source size can be assumed to be significantly slower than 0.01~c which is linked to the
corresponding adiabatic expansion time scales.

\subsubsection{Conclusion for the modelling approach}
\label{subsubsec:concmodel}
From this it can be concluded that the intrinsic flux density evolution of 
X-ray luminous components is dominated by the time scales of the relevant 
sub-mm peaking synchrotron components.
This then implies that rapid flux density variations in the X-ray domain are
more likely a result of relativistic effects like boosting and lensing 
rather than synchrotron cooling due to high energetic electrons.
This is supported by the fact that the time scales for the boosting and lensing events
are very short (in the few minutes to 10 minute range - depending on the geometry)
compared to those on which the
underlying emission is varying. Also (again depending on the geometry) the amplification
factors of these relativistic effects can easily be several 10 to 100.
In addition, the flare lengths indicate that the flares themselves originate 
close to the black hole (Baganoff et al. 2001, 2003)
making these relativistic modulations also frequent and relevant.
This scenario justifies that we model the flare profiles with an orbiting spot model.
This model is a generalized surrogate-model to characterize the behavior of (in this 
case) X-ray emitting matter in the gravitational field of a supermassive black hole.

Of course, we only consider the variable non-thermal part of the SgrA* emission.
This source is embedded in an extended non-variable Bremsstrahlung component
(Baganoff et al. 2001, 2003) and during times the source is not flaring,
the non-thermal flux drops well below the Bremsstrahlung flux level.
This constant component is therefore removed before we fit the flares.
Also $\beta_{X-ray}$ is significantly smaller than $\beta_{NIR}$  (see above),
hence, brighter flares in the X-ray domain are much more pronounced.
Restricting the modelling to the brightest flares, therefore essentially 
completely avoids the risk of overlapping flare events, such that we 
truly model only one flare at a time.
\\
\\
This also supports the choice of spots on circular orbits as a 
surrogate model for radiating matter close to the SMBH
as highly elliptical orbits due to in-falling matter are 
probably strongly suppressed.
In a viscous environment with multiple gaseous clouds  
(as expected for 'central mid-planes' resulting 
from magneto-hydrodynamic accretion models),
clouds on crossing orbits can be excluded as their collisions
are highly dissipative.
As (semi-)stable trajectories in such an environment circular 
orbits are preferred.
Hence, cases of freely in-falling matter on highly elliptical trajectories 
close to the SMBH are probably rare and will not dominate the 
observed flare cases.
But even if they occur and even if the viscosity is small one can state 
the following:
Is the distance of the spot equal or shorter than the semilatus rectum
for highly elliptical trajectories and for cases for which the line 
of sight passes close to the SMBH so that the relativistic effects 
are relevant,
the difference in distance to the SMBH is between a factor 1 to 2,
compared to the circular case. The velocity is less than 
$\sqrt{2}$ higher if the spot is at its priapse.
In proper motion this difference is largely compensated for 
if the spot is observed close to the semilatus rectum position.
In line of sight velocity this factor comes fully in only if the 
long axis is almost perpendicular to the line of sight.
On the one hand this shows that the expected differences in the 
light curve shapes are well covered by differences for light 
curves shapes for circular orbits in our experimental setup.
On the other hand the length of the observed light curve $T_{obs}$ 
effected by relativistic effects
is only a factor of $\lesssim$$\sqrt{2}$ longer than in the circular case.
For all other orientations ($\le$50\% in three dimensions) 
of the highly eccentric trajectories
the lensing case becomes quickly unlikely and the boosting becomes too small
to effect the spot luminosity. 
Hence, while highly eccentric orbits for viscous magneto-hydrodynamic 
accretion are not likely anyhow, these cases are also not likely to
play a dominant role if they occurred in the immediate vicinity 
of a SMBH.

However, in general, and not necessarily in the current context of 
hot spots in the framework low luminosity temporary 
accretion disks, eccentric accretion flows may play an important role
on much longer time scales.
Eccentric accretion flows are naturally formed after tidal disruption
events. The initial eccentricity of tidal debris scales with the ratio
of the stellar mass (M$_{star}$) to the black hole mass ($M_{bh}$)
approximately as 1-2(M$_{star}$/M$_{bh}$)$^{1/3}$, approaching unity for
smaller ratios (Svirski et al. 2017). Hot spots in the eccentric
tidal streamers could be formed as shocks as a result of the
supersonic collision of the tidal debris with the initial streamer at
the apo-center of the orbit. However, such events are rather rare for
Sgr A*, since statistically only one disruption of a star per 10$^4$
years is expected (Phinney 1989).
Other hypothetical source of eccentric synchrotron 'hot spots' would be young 
neutron stars, since they wouldn't be tidally disrupted even at ISCO. 
So for some time they could orbit SMBH on precessing eccentric orbits until they 
would plunge through the horizon due to the gravitational-wave losses. 
This is similar to the original idea of hot spots being stars as studied first 
by Cunningham and Bardeen (1973). The full relativistic treatment of an eccentric 
hot spot is given by Bao, Hadrava, \& Ostgaard (1994) - they compare the 
light curves for zero eccentricity and e=0.5 with a similar set-up as used by 
Karssen et al. (2017).

\subsubsection{The $\gamma$-ray domain}
\label{subsubsec:gamma}
While the overall SgrA complex has been detected by ground-based Cherenkov Telescopes 
(e.g. Aharonian et al. 2004, Kosack et al. 2004, Albert et al. 2006) no unique SgrA* counterpart has 
been reported in the $\gamma$-ray domain.
Ahnen et al. (2017) present results of a monitoring campaign of the Galactic center (GC) using the 
MAGIC Imaging Atmospheric Cherenkov Telescopes between 2012-2015.
This was done with the goal of detecting high-energy flare emission above 100~GeV under the 
assumption that the passage of the Dusty S-cluster Object leads to enhanced accretion events 
during and after its SgrA* periapse in mid-2014 (e.g. Valencia et al. 2015).
Previous results on the $\gamma$-ray spectrum of SgrA* were confirmed but no significant 
variability  or excess emission was detected.
Photons detected from SgrA* in this energy domain would most likely be comptonization NIR of X-ray photons.
Future observation with the Cherenkov Telescope Array (CTA) (https://www.cta-observatory.org/)
may reveal more information on the  $\gamma$-ray spectrum of SgrA* or the central region in general.

\section{Summary}

SgrA* can be considered as an extremely low luminosity LLAGN. Previous
activities may have been higher than today (Fig.~\ref{fig2}).
For LLAGN there is an indication for a mass dependent positioning 
or evolution from normal star bursting galaxies 
to Seyferts and to LINERS, or even via Seyferts to LINERS
(Fig.~\ref{fig2}).
There are several indications that star formation on all mass scales
may take place within the central stellar cluster or even close to the 
supermassive black hole (see section \ref{sec:Hintsforstarformation}).

From the orbit of S2 there are close hints at the fact that especially the
the very
central S-cluster stars are on relativistic orbits around SgrA*.
The accretion process and the corresponding emission from SgrA*
have characteristic signatures (spectral index, degree of variability)
in all wavelength domains from the
radio via the infrared to the X-ray regime.
These characteristic signatures may in part 
(at least for the brighter flares)  also include the
possible flare shapes.
Here, the emission of orbiting hot-spots is modulated by 
boosting and lensing  on his orbit around the SMBH.
The dominant radiation mechanism for flares in the sub-mm to X-ray domain is
synchrotron mechanism in combination with a comptonization of the
corresponding photons.

\section*{Acknowledgements}
We received funding
from the European Union Seventh Framework Program
(FP7/2013-2017) under grant agreement no 312789 - Strong
gravity: Probing Strong Gravity by Black Holes Across the
Range of Masses. 
This work was supported in part by the
Deutsche Forschungsgemeinschaft (DFG) via the Cologne
Bonn Graduate School (BCGS), the Max Planck Society
through the International Max Planck Research School
(IMPRS) for Astronomy and Astrophysics, as well as special
funds through the University of Cologne and
SFB 956 - Conditions and Impact of Star Formation. M. Zajacek, M. Parsa and
B. Shahzamanian are members of the IMPRS. Part of this
work was supported by fruitful discussions with members of
the European Union funded COST Action MP0905: Black
Holes in a Violent Universe and the Czech Science Foundation
- DFG collaboration (No. 13-00070J).

\end{document}